\documentclass[iop]{emulateapj}
\usepackage{natbib}
\slugcomment{Draft Version November 18, 2014 }

\shorttitle{Outward Motion of Porous Dust Aggregates in Protoplanetary Disks}
\shortauthors{Tazaki and Nomura}

\begin{document}


\title{Outward Motion of Porous Dust Aggregates by Stellar Radiation Pressure\\ in Protoplanetary Disks}



\author{RYO TAZAKI\altaffilmark{1}}
\affil{Department of Astronomy, Graduate School of Science, Kyoto University, Kitashirakawa-Oiwake-cho, Sakyo-ku, Kyoto 606-8502, Japan}
\and
\author{HIDEKO NOMURA}
\affil{Department of Earth and Planetary Sciences, Tokyo Institute of Technology, 2-12-1 Ookayama, Meguro-ku, Tokyo 152-8551, Japan}


\altaffiltext{1}{rtazaki@kusastro.kyoto-u.ac.jp}


\begin{abstract}
We study the dust motion at the surface layer of protoplanetary disks. 
Dust grains in surface layer migrate outward due to angular momentum transport via gas-drag force induced by the stellar radiation pressure. 
In this study, we calculate mass flux of the outward motion of compact grains and porous dust aggregates by the radiation pressure.
The radiation pressure force for porous dust aggregates is calculated using the T-Matrix Method for the Clusters of Spheres.
First, we confirm that porous dust aggregates are forced by strong radiation pressure even if they grow to be larger aggregates in contrast to homogeneous and spherical compact grains to which efficiency of radiation pressure becomes lower when their sizes increase.
In addition, we find that the outward mass flux of porous dust aggregates 
with monomer size of 0.1 $\mu$m is larger 
than that of compact grains by an order of magnitude at the disk radius of 1 AU, 
when their sizes are several microns.
This implies that  large compact grains like calcium-aluminum rich inclusions (CAIs) are hardly transported to outer region by stellar radiation pressure, whereas porous dust aggregates like chondritic-porous interplanetary dust particles (CP-IDPs) are efficiently transported to comet formation region.
Crystalline silicates are possibly transported in porous dust aggregates by stellar radiation pressure from inner hot region to outer cold cometary region in the protosolar nebula.
\end{abstract}





\section{Introduction}
Protoplanetary disks (hereafter PPDs) are formed around the protostar and are thought to be the site of on-going planet formation. 
Present-day solar system provides us useful insights to the planet formation 
theory. Among them, there have been various debates about the origin of crystalline silicates in comets. 
Spectroscopic observations of comets have revealed that comets contain substantial amount of crystalline silicates. 
The mass ratio of crystalline silicate to total (amorphous+crystal) silicate has wide variety, such that 70\% for C/2001 Q4 \citep{wooden04,ootsubo07}, $30\sim35$\% for 73P/S-W3 \citep{harker11, sitko11} and 14\% for C/2007 N3 \citep{woodward11}. 
Crystallinity is a smoking gun of thermal processing, however, comets are thought to be formed at outer cold region of 
PPDs where the thermal processing is not likely to occur. 

In general, crystalline silicate is formed by (i) annealing of amorphous silicate and (ii) direct condensation from gas phase \citep[e.g.,][]{gail10}.
Annealing of amorphous silicate is a process of re-arrangement of lattice structure using the external thermal energy. Laboratory experiment shows
 that the crystallization timescale is $10^6$ yrs for $T=800$ K and 1 yrs for $T=1000$ K \citep[e.g.,][]{murata09}. 
In the case of vaporization and subsequent gas phase condensation, required gas temperature is at least, $1380\sim1310$ K at $10^{-4}$ bar \citep{lodders03}.
This is rather high temperature compared with that of comet formation region, tens of kelvins.

Since crystalline silicates are depleted in the interstellar medium, which is due to cosmic-ray hit or particle bombardment \citep{kemper04, kemper05}, crystalline silicates must be formed in the PPDs. Indeed, many observations showed the presence of crystalline silicate in the PPDs around T-Tauri stars \citep{bouwman08,olofsson09,watson09}, Herbig Ae/Be stars \citep{juhasz10} and young brown dwarf stars \citep{riaz12}. 
However, the formation location for the crystalline silicate is still being debated.
One possibility is annealing or condensation at inner region of PPDs where the viscous heating yields high temperature environment \citep{gail04}. 
Another possibility is an in situ annealing at larger distance from the central star by shock wave \citep{desch02, harker02} or annealing inside the clump 
fragmented from massive disk \citep{vorobyov11}. 
The other possibility is an episodic heating by intense radiation from central star in outburst phase. Such episodic heating event could anneal the amorphous silicate in the surface layer of PPDs \citep{abraham09, juhasz12}.

Chondritic porous interplanetary dust particles (CP-IDPs) can make constraint on the origin of crystalline silicates in comets, since they are considered as a cometary origin \citep[e.g.,][]{messenger14}.
CP-IDPs are highly porous aggregates of sub-micron dust grains mainly made of crystalline silicate (enstatite and forsterite) and amorphous silicate called GEMS (glass with embedded metal and sulfides).
Crystalline morphology found in enstatite in CP-IDPs such as enstatite whisker or platelet
suggests the enstatite is formed via direct condensation from gas phase \citep{bradley83}.
The formation of enstatite via condensation at inner region is also suggested by both theoretical calculation based on equilibrium condensation model in PPDs \citep{gail04} and observations \citep{bouwman08, juhasz10}. 
 Moreover, although the origin of GEMS is still controversial, recent studies suggested GEMS can be formed via non-equilibrium condensation
 from gas-phase \citep{keller11, matsuno14}. Therefore, it is possible to form most of components in CP-IDPs at inner region of PPDs.

While processed materials in CP-IDPs are formed at inner region (less than $1$ AU), comets are formed at outer region of PPDs (more than $30$ AU, although it is still unclear). Thus it is suggested that in the early phase of PPDs, there was a large scale radial mixing which connect the inner and the outer regions of protoplanetary disks.
In order to explain this large scale radial mixing, many models have been proposed so far.
One of the plausible ideas for radial mixing is a diffusion of dust grains induced by turbulent gas
\citep{gail01, gail04, bockelee-morvan02, cuzzi03, hughes10}.
Gas in the protoplanetary disks is turbulent due to the magneto-rotational instability \citep{balbus91} and the
dust grain whose size is small enough to be coupled to gas is able to diffuse to outer part to some extent. However, turbulence also induces accretion toward the central star
and it prevents grains from spreading outer part of the disk. 
To transport large amount of refractory grains to outer part of PPDs against the accretion flow, evolution of initially compact disk is important \citep{bockelee-morvan02, cuzzi03}.
Meridional flow model is another model for radial mixing \citep{keller04, ciesla07, ciesla09, hughes10}.
This model is based on midplane outward flow suggested by \citet{urpin84} and \citet{takeuchi02} in the two-dimensional viscous disk.
However global MHD simulation suggested the meridional flow does not appear in MRI-turbulent disks \citep{fromang11}.
Another model is the surface outflow of the dust grain by the radiation pressure \citep{takeuchi03, vinkovic09}.
\citet{takeuchi03} found that the dust grain in the optically thin layer are exposed to the stellar radiation pressure and it leads to the outward motion
 through the angular momentum exchange between dust grain and gas.

Here, we model the outward motion of dust grains in the surface layer of protoplanetary disks by the stellar radiation pressure based on the model of \citet{takeuchi03}. To mimic CP-IDPs, we consider porous aggregates of small spherical grains, called monomers.
Although previous studies assumed homogeneous and spherical compact grain, it is presumed that the radiation pressure to the porous dust aggregates is more efficient than the compact sphere with equivalent radius so that the surface outflow of porous dust aggregates works more efficiently than the compact grain. 
One of the most important parts of this work is a calculation of optical properties of porous dust aggregates. 
In many studies, optical properties of porous dust aggregates
are calculated using simply approximated methods, such as the
effective medium theory \citep[e.g.,][]{chylek00}. However, this method is applicable to the particle whose size
 is smaller than the wavelength. 
The peak wavelength of black body radiation from
a typical T-Tauri star or protostar
is sub-micron and it is roughly comparable to 
the size of monomer of porous dust aggregates. Therefore, we should use a
scheme that has an accuracy in such wavelength. In this work we use
the T-Matrix Method that is a one of most precise methods for calculating optical properties of porous dust aggregates (see Sect. 2.3.2 for more details). 

The plan of this paper is as follows. 
In section 2, we introduce the model of gas/dust disk and optical properties of grains.
In section 3, we show the result of optical properties of porous aggregates and dynamics of dust grain at the surface layer of PPDs,
 and discuss the result. Finally, in section 4, we state the implication to the origin of cometary grains.

\section{Model}

In this paper, we adopt cylindrical coordinate $(r,\phi,z)$ and assumed
an axisymmetric and geometrically thin disk. 
\subsection{Gas Disk}
We assume that gas is in hydrostatic equilibrium in the vertical
direction, 
\begin{equation}
	-\frac{GM_{*}z}{(r^2+z^2)^{3/2}}-\frac{1}{\rho_g}\frac{\partial{P_g}}{\partial{z}}=0,
\end{equation}
where $G$ is gravitational constant and $M_*$ is the mass of the central
star, and $\rho_g$ and $P_g$ are the density and pressure of gas,
respectively. 
We neglect here the self-gravity of disk. Suppose isothermal in z-direction, 
the gas density is given by
\begin{equation}
	\rho_g(r,z)=\rho_{g}(r,0)\exp\left(-\frac{z^2}{2h_g^2}\right).
\end{equation}
${h}_g$ is the gas scale height defined as
 \begin{equation}
 h_g=c_s/\Omega_{K,{\rm mid}}, \label{eq:hg}
\end{equation}
where $c_s$ is the sound speed, 
and $\Omega_{K,{\rm mid}}=
({GM_*}/{r^3})^{1/2}$ is the
angular frequency at the midplane of Keplerian disks. 
The force balance in radial direction respected to gas particle is
\begin{equation}
	r\Omega_g^2-\frac{GM_{*}r}{(r^2+z^2)^{3/2}}-\frac{1}{\rho_g}\frac{\partial{P_g}}{\partial{r}}=0.
\end{equation}
If we define 
$\eta$ as a ratio of gas pressure and stellar gravity, 
$\eta=-(r\Omega_K^2\rho_g)^{-1}\partial P_g/\partial r$, then
\begin{equation}
	\Omega_g=\Omega_{K,{\rm mid}}\left(1-\eta\right)^{1/2}, \label{eq:Omgg}
\end{equation}
where we neglect higher order than $(z/r)^2$.

For simplicity, the radial profiles of 
the gas density and temperature
are assumed to have power-law forms, such that
\begin{eqnarray}
\rho_g(r,0)&=&\rho_{g,0}\left(\frac{r}{1\rm{AU}}\right)^p,\\
T(r) &=& T_0 \left(\frac{r}{1\rm{AU}}\right)^ q, 
\end{eqnarray}
where 
$T$ is the temperature. 
Then the gas scale height in Eq. (\ref{eq:hg}) becomes
\begin{equation}
h_g=h_0\left(\frac{r}{1\rm{AU}}\right)^{(q+3)/2}.
\end{equation}
The surface density of gas disk is reduced to
\begin{eqnarray}
\Sigma_g(r) &=& \int_{-\infty}^{\infty}\rho_g(r,z)dz \nonumber \\
&=& \sqrt{2\pi}\rho_{g,0}h_0\left(\frac{r}{1\rm{AU}}\right)^{p+(q+3)/2}.
\end{eqnarray}
Also, $\eta$ can be written in the following form \citep{takeuchi02}
\begin{eqnarray}
\eta=-\left(\frac{h_g}{r}\right)^2\left(p+q+\frac{q+3}{2}\frac{z^2}{h_g^2}\right).
\end{eqnarray}

In this paper, we adopt $p=-2.25$, $q=-0.5$, $\rho_{g,0}=2.83\times10^{-10}$ g/cm$^3$, $h_0 = 3.33 \times 10^{-2}$ AU and $T_0=278$ K. Then the disk mass within $40$ AU is 0.01 $M_{\odot}$ and this is comparable to Minimum Mass Solar Nebula \citep{hayashi85}. 
Also, in this case $\eta$ is positive near the midplane, 
while it becomes negative
at the surface layer ($z\ga 1.5h_g$). 
Therefore, the gas disk rotates with sub-Keplerian velocity near the
midplane and with super-Keplerian velocity at the surface layer.

Gas disk accretes onto star due to angular momentum transport induced by the MHD turbulence.
The conservation of mass and angular momentum are reduced to
 \begin{equation}
 v_{g,r}=-\frac{3\nu}{r}\frac{\partial{\ln(\Sigma_g\nu{r}^{1/2})}}{\partial{\ln{r}}}. \label{eq:vgr}
 \end{equation}
For the model of turbulent viscosity, we assume the $\alpha$-model \citep{shakura73},
\begin{equation}
\nu=\alpha{c_s}h_g, \label{eq:vis}
\end{equation}
 where $\alpha$ is a non dimensional quantity that indicates the
strength of turbulent viscosity. We adopt $\alpha=10^{-3}$ and assume it is constant in vertical direction
 in this paper.
 
\subsection{Dust Disk}
 For dust grains smaller than the mean free path of gas molecules, the
Epstein's law is applicable. The mean free path of gas molecules in the disks is given by
\begin{eqnarray}
 \lambda_{\rm mfp}=\frac{\mu{m_{\rm H}}}{\sigma_{\rm mol}\rho_g}
 =6.9\left(\frac{r}{1\rm AU}\right)^{2.25} \rm{cm},
\end{eqnarray}
where $\mu=2.34$ is the mean molecular mass, $m_{\rm H}$ is the mass of hydrogen atom,
 $\sigma_{\rm mol}=2\times10^{-15}$ cm$^2$  is collision cross-section of gas molecules. The grain radius we treat in this paper is always smaller than the mean free path.
The Epstein's law of gas drag force is given by 
\begin{equation}
F_{\rm{drag}} = - \frac{4}{3}\rho_g\sigma{v}_t v,
\end{equation}
where $v_t=\sqrt{8/\pi}c_s$ is the thermal velocity of gas, $\sigma$ is the geometrical
cross-section of dust grain and $v$ is the relative velocity between gas
and dust. The timescale of the gas drag force can be estimated as
\begin{equation}
t_s = \frac{mv}{|F_{\rm{drag}}|} = \frac{3}{4}\frac{m}{\sigma}\frac{1}{\rho_gv_t},
\end{equation}
where $m$ is the mass of 
a dust particle. 
It represents the typical timescale in which dust grains lose their momentum by the gas drag force. This timescale is called ``stopping time" and it is convenient to use the stopping time normalized by the dynamical time, $T_s \equiv t_s\Omega_{K,\rm{mid}}$. In the case of $T_s \ll 1$, dust grains stop so quickly  relative to gas that the dust grains move just like gas (well coupled). If $T_s \gg 1$, the gas drag force is ineffective therefore the dust grains move independent of gas (decoupled). 

If we assume that the distribution of dust particles in the vertical
direction is controlled by turbulent mixing of gas, the density
distribution of dust grains is derived as
\begin{equation}
\rho_d(r,z) = \rho_d(r,0)\exp\left[-\frac{z^2}{2h_g^2}-\frac{T_{s\rm{,mid}}}{\alpha}\left(\exp\frac{z^2}{2h_g^{2}}-1\right)\right],
\end{equation}
where $T_{s\rm{,mid}}$ is the non-dimensional stopping time at midplane \citep{takeuchi03}.
Note that the density distribution of dust grains is deviated from the Gaussian distribution
at high altitude ($z\geq h_g$) since dust grains become decoupled from gas.
$\alpha$ in Eq.(16) is a non-dimensional parameter for the turbulent diffusion
coefficient and we simply assume that $\alpha$ is the same as the parameter for
the turbulent viscosity in Eq. (\ref{eq:vis}) in this paper. 
The dust density at the midplane, $\rho_d(r,0)$, is derived so that the
dust surface density,  
\begin{equation}
\Sigma_d(r)=\int_{-\infty}^{\infty}\rho_d(r,z)dz,
\end{equation}
satisfies $\Sigma_d(r)=f_{\rm{dust}}\Sigma_g(r)$, where $f_{\rm{dust}}$
is the dust-to-gas mass ratio and we adopt $f_{\rm{dust}}=0.01$ in this paper.

The equation of motion of dust grains is
\begin{eqnarray}
&&\frac{dv_{d,r}}{dt}=\frac{v_{d,\phi}^2}{r}-(1-\beta)\Omega_{K, \rm{mid}}^2r-\frac{v_{d,r}-v_{g,r}}{t_s}, \label{eq:eqdr}\\
&&\frac{d}{dt}(rv_{d,\phi})=-r\frac{v_{d,\phi}-v_{g,\phi}}{t_s}, \label{eq:eqdphi}
\end{eqnarray}
where $v_{d,r}$ and $v_{d,\phi}$ are the radial and azimuthal velocities of dust grains, respectively, and $v_{g,\phi}$ is the azimuthal velocity of gas. 
$\beta$ is a ratio of radiation pressure and stellar gravity.
We neglect the effect of Poynting-Robertson drag in both Eq. (\ref{eq:eqdr}) and Eq. (\ref{eq:eqdphi}) 
since it is negligible compared to the gas drag force 
in gas-rich protoplanetary disks
\citep{takeuchi01}. Assuming $v_{d,\phi}\sim v_{g,\phi}\sim v_{K}$ and steady state, and solving Eq.(\ref{eq:Omgg}), Eq. (\ref{eq:eqdr}) and Eq. (\ref{eq:eqdphi}), we get
\begin{equation}
	v_{d,r}=\frac{v_{g,r}T_{s}^{-1}+(\beta-\eta){v}_{K}}{T_s+T_s^{-1}} \label{eq:vel_dr} ,
\end{equation}
where $v_K=r\Omega_{K,\rm{mid}}$ is the Keplerian velocity.
Eq.(\ref{eq:vel_dr}) shows that the dust radial velocity is coincident with gas accretion velocity when the dust grains are well coupled to gas ($T_s \ll 1$) and  the dust grains with large $\beta$ compared to $\eta$ can move outward when they are weakly decoupled from gas ($T_s \sim 1$).
When the gas drag force is negligible, Eq.(18) shows that the rotation frequency of dust particles becomes
\begin{equation}
\Omega_d=\Omega_{K, {\rm mid}}(1-\beta)^{1/2}, \label{eq:Omgd}
\end{equation} 
and the dust particles orbit around the central star with
sub-Keplerian frequency due to radiation pressure in surface layer. 

In optically thin surface layer where the stellar radiation is not
absorbed by dust grains, the ratio is reduced to \citep{burns79}
\begin{eqnarray}
\beta_{0} \equiv \frac{F_{\rm{RP},0}}{F_{\rm{grav}}}
 = K\left(\frac{\sigma}{m}\right)\int_0^\infty Q_{\rm{RP}}(x,\mathfrak{m})B_{\lambda}(T_{*}) d\lambda \label{eq:beta},
 \end{eqnarray}
 where $K=\pi R_*^2/(GM_*c)$ is a constant, $R_*$ is the
radius of the central star, and
$B_{\lambda}(T_{*})$ is the Planck function at the effective
temperature of the central star, $T_*$, and the wavelength of incident
radiation, $\lambda$. We assume $M_{*}=M_{\odot}$ and
$T_*=5778$ K, 
which is plausible value for the stellar effective temperature in outburst phase\ \citep[e.g.,][]{juhasz12}. 
$Q_{\rm{RP}}(x,\mathfrak{m})$ is the
efficiency of radiation pressure, and $x$ and $\mathfrak{m}$ are
the size parameter and the refractive index of dust grains, respectively
(see Sect. 2.3 for more details).
Both of the gravity and radiation pressure inversely proportional to the square
of disk radius, hence $\beta_{0}$ does not depend on the distance from the central star.

The presence of dust grains makes the protoplanetary disk optically thick.
The irradiation from the central star, and then the radiation pressure is reduced
by dust absorption as
$F_{\rm RP}=F_{{\rm RP},0}\exp(-\tau)$, and $\beta$ becomes
\begin{equation}
\beta=\beta_0\exp[-\tau(r,z)].
\end{equation}
We note that contribution of stellar radiation scattered on dust
grains is simply neglected in this study.
We calculate the optical depth from the central star as
\begin{equation}
\tau(r,z)=\int_{r_{\rm{in}}}^{r}\frac{\kappa_{\rm{abs, disk}}}{f_{dust}}\rho_d(r',z')\sqrt{1+\frac{z^{\prime 2}}{r^{\prime 2}}}dr', \label{eq:tau}
\end{equation}
where $r_{\rm{in}}$ is the inner radius of disk, which is assumed to be $r_{\rm{in}}=0.1$ AU, and $\kappa_{\rm{abs, disk}}$ is the
mass absorption opacity and we simply adopt the typical value at visible wavelength,
$\kappa_{\rm{abs, disk}}=100$ g/cm$^2$.
\begin{table*}[htbp]
\begin{center}
\caption{The characteristic radius$s_c$ and $\beta$-value of porous aggregates with
 monomer number of $N$}
\begin{tabular}{|l|lllllllll|}
\hline
$N$ & 1 & 8 & 16 & 32 & 64 &128 & 256 & 512 & 1024\\
\hline\hline
$s_c$ & 0.1 & 0.33 & 0.46 & 0.74 & 1.04 &1.41 & 2.34 & 3.09 & 4.37\\
$\beta$ & 0.73 & 0.69 & 0.68 & 0.67 & 0.66 &0.65 & 0.64 & 0.64 & 0.62\\
\hline
\end{tabular}
\label{tab:rc}
\end{center}
\end{table*}

\subsection{Dust Model}
We deal with porous aggregates composed of number of spherical monomers.
BCCA (Ballistic-Cluster-Cluster Aggregation) and
BPCA (Ballistic-Particle-Cluster-Aggregation) are widely used as models for porous aggregates.
The degree of fluffiness can be described by the fractal dimension $D$ defined by
\begin{equation}
N\propto\left(\frac{s_c}{s_0}\right)^{D}
\end{equation}
where $N$ is the monomer number$, s_0$ is the monomer radius and $s_c$ is the characteristic radius of porous aggregates.
BCCA and BPCA show $D\sim2.0$ and 3.0, respectively. 
The characteristic radius of porous aggregates is given by
\begin{equation}
s_c^2 = \frac{3}{5}s_g^2 ,\ s_g=\left[\frac{1}{2N^2} \sum_{i}^{N} \sum_{j}^{N}({\mathbf s_i}-{\mathbf s_j})^2\right]^{1/2},
\end{equation}
where ${\mathbf s_i}$, ${\mathbf s_j}$ are the position vector
of the $i$-th and $j$-th monomers \citep[e.g.,][]{mukai92}.
In this 
paper, three types of dust grain, 
a monomer, compact grains and porous aggregates, are considered.
We define the filling factor of porous aggregates as,
\begin{equation}
f=N\left(\frac{s_0}{s_c}\right)^3. \label{eq:filling}
\end{equation}

Porous aggregates are assumed to be BCCA 
with $8$, $16$, $32$, $64$, $128$, $256$, $512$ and $1024$ monomers.
The monomer radius of $s_0=0.1$ $\mu$m is adopted, which is a typical size of GEMS or crystalline enstatite. Dust physical density of olivine, $\rho_s=3.3$ g/cm$^3$ \citep{draine84}, is used.
  
\subsubsection{Optical Properties of Dust Grains}
Optical properties, or the scattering and absorption by a particle,
vary drastically 
depending on it's size, shape and chemical composition.
The dependence on size and wavelength can be scaled by the size parameter,
\begin{equation}
x=\rm{k}s=\frac{2\pi s}{\lambda},
\end{equation}
where $s$ is the size of particle, $\rm{k}$ is the wave number and $\lambda$ is an incident wavelength.
The other factor that controls optical properties is refractive index of particle.
It is convenient to use the complex representation of refractive index,
$\mathfrak{m}=n+ik$ where $n$ and $k$ is the real part and
imaginary part of refractive index, respectively.
Note that in this paper, the symbol $\rm{k}$ denotes the wavenumber while $k$ represents the imaginary part of complex refractive index.
Real part of $\mathfrak{m}$ determines the phase velocity in the medium and imaginary part represents the absorption by the medium. 
In this paper, the chemical composition of dust grain is assumed to
be the astronomical silicate \citep[e.g.,][]{draine84,weigartner01,li01}.
It should be noted that astronomical silicates are amorphous, and in general optical constants differ between amorphous and crystal. An anisotropic crystal, such as an enstatite whisker, has different optical properties respect to each crystalline axis, thus the Mie Theory, and then the T-Matrix Method,  are not applicable without some approximation. Therefore, in this paper, we simply assume homogeneous, isotropic amorphous silicate as a monomer.

Once we know the size parameter and complex refractive index, we can
calculate absorption/scattering cross-section. 
We define the efficiency for absorption and scattering as $Q_{\rm{abs}}=C_{\rm{abs}}/\sigma$ and $Q_{\rm{sca}}=C_{\rm{sca}}/\sigma$ where
 $C_{\rm abs/sca}$ is the absorption/scattering cross-section
and $\sigma$ is the geometrical cross-section of the dust particle. 
Using the absorption and scattering efficiency, we can describe the
radiation pressure efficiency as
\begin{equation}
 Q_{\rm{RP}}=Q_{\rm{abs}}+(1-g)Q_{\rm{sca}}, \label{eq:qrp}
\end{equation}
where $g$ is the asymmetry parameter defined by $g=<\cos\theta>$.
Here,
$\theta$ is the scattering angle. Asymmetry parameter represents the degree of forward scattering.
For the longer wavelength ($x\ll 1$), $g=0$
due to Rayleigh scattering and then
$Q_{\rm{RP}}\sim Q_{\rm{ext}}$.
Also in the Rayleigh regime, extinction efficiency is dominated by absorption efficiency. Eventually, $Q_{\rm{RP}}\sim Q_{\rm abs}$.
If the size parameter is large ($x\gg 1$), asymmetry parameter approaches to unity and then $Q_{\rm{RP}} \sim Q_{\rm{abs}}$.
The scattering plays an important role for radiation pressure when $x\sim1$.

\subsubsection{Optical Properties of Porous Aggregates}
We can calculate these optical properties using the Mie Theory
if the grain is homogeneous, isotropic and spherical \citep{bohren83}.
However, in the case of inhomogeneous aggregates this semi-analytical method is not applicable.
In order to calculate the optical properties of porous aggregates, we need numerical methods.

One of the simplest ways is using the Effective Medium Theory (EMT) \citep[e.g.,][]{chylek00}. 
This theory deals with porous aggregates as a sphere with effective optical constants.
The formulation of the EMT assumes that (i) incident light is static fields, which is approximately satisfied unless the inverse of the frequency of incident lights falls the timescale of polarization of monomer (Rayleigh limits) and (ii) neglecting the interaction between monomer's scattered fields.
Therefore, in principle, the EMT cannot handle the shorter wavelength domain where the higher order of Lorentz-Mie coefficients becomes important. 

Another method commonly used is the Discrete Dipole Approximation (DDA) \citep{purcell73, draine94}.
The advantage of DDA is applicability to arbitrary shaped, inhomogeneous and anisotropic particle. 
If monomer is divided into $N_{\rm d}$-dipoles having the dipole polarizability determined via radiative reaction corrections \citep{draine88} and lattice dispersion relation \citep{draine93}, 
the DDA can treat porous aggregates and yields correct results, when $N_{\rm d}$ is taken to be large enough to converge the calculation.
However, with increasing $N_{\rm d}$, the computation of the DDA requires huge computing memory and long computing time. 
One way to reduce numerical costs is to replace spherical monomers as single dipoles \citep[e.g.,][]{okamoto98}, however, this method only applicable for $x_m<1$.

The T-Matrix Method for the Clusters of Spheres (TMM) is also widely accepted method \citep[e.g.,][]{mishchenko96,okada08}. 
This is one of the most rigorous methods of calculating
optical properties of porous aggregates
 using the multi-sphere superposition method. 
The TMM is applicable to $x_{\rm v}<100$ where $x_{\rm v}$ is a size parameter for the volume equivalent sphere \citep[e.g.,][]{mishchenko00}.
 
The peak wavelength of the blackbody radiation at $T_*=5778$ K
is $\lambda_{\rm peak}\sim 0.5$ $\mu$m so that, at the peak wavelength, size parameter for the monomer is $x_m\sim 1.25$ and volume equivalent size parameter is $x_{\rm v}\sim{12.6}$ when the monomer size is $0.1$ $\mu$m and monomer number is $N=1024$. 
Therefore, we can use the T-Matrix Method.
We use the code by \citet{okada08}, with which faster computation speed is available by adopting the 
quasi-orientation averaging\footnote{This is an open code and one can download it from following website. \url{http://www.iup.uni-bremen.de/\textasciitilde{alexk}/page26.html}}. 
In our calculation, we averaged optical properties over 30 orientations for each aggregates.

\section{Results}
First, we report the difference of $\beta$-value, the ratio of the radiation pressure to stellar gravity, between porous aggregates and compact grains. 
And next using the result, we calculate the radial velocity of dust grain at the surface layer and estimate the outward mass flux.
\subsection{Radiation Pressure}
\begin{figure}[t]
\begin{center}
\includegraphics[height=6.5cm]{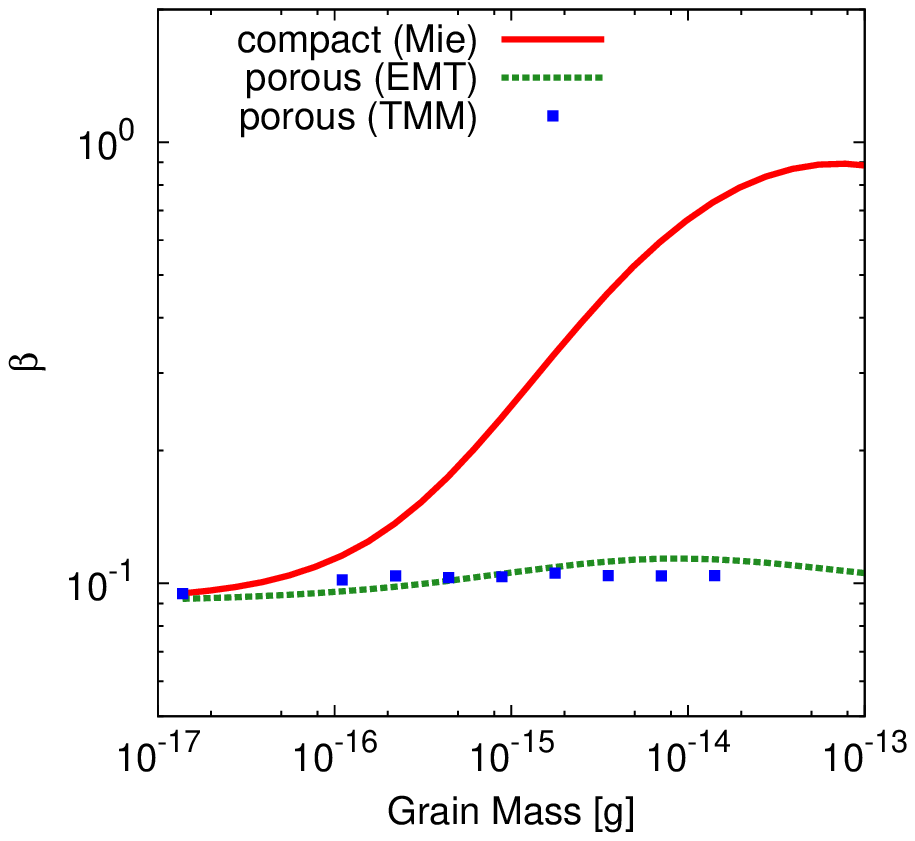}
\includegraphics[height=6.5cm]{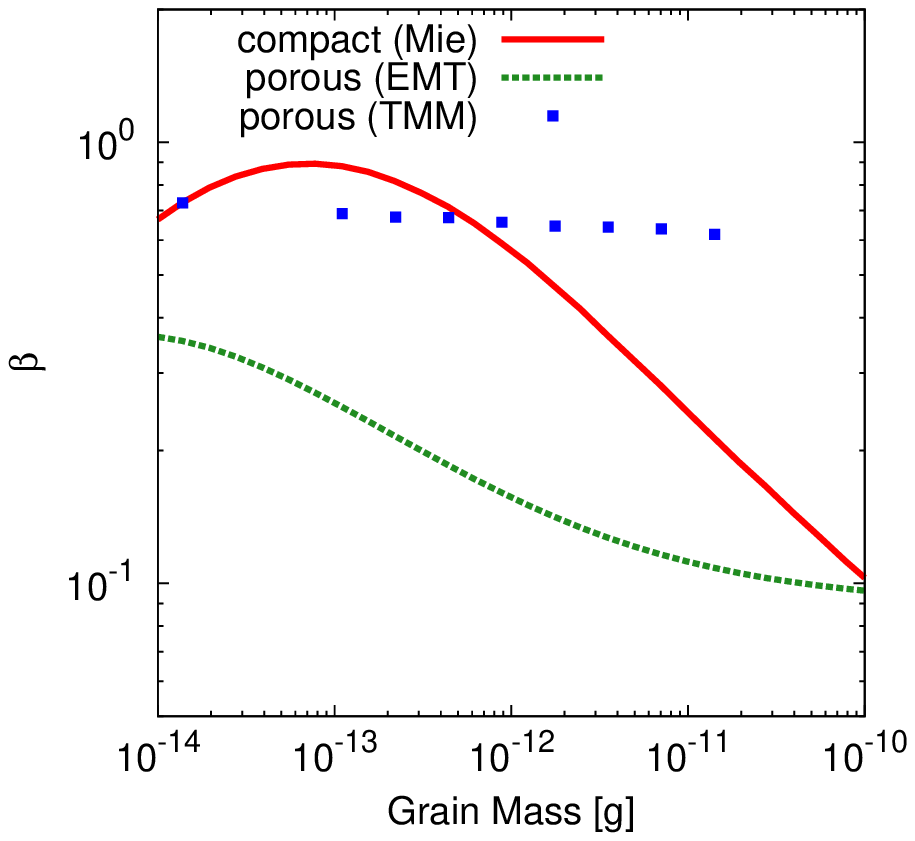}
\caption{$\beta$ for the compact silicate grain ({\it red lines}) and porous aggregates (BCCA) with N=1 (monomer), 8, 16, 32, 64, 128, 256, 512 and 1024 obtained by the T-Matrix Method ({\it blue squares}).
Top panel is the case for $s_0=0.01$ $\mu$m and bottom panel is for $s_0=0.1$ $\mu$m. 
$\beta$ for porous aggregates are almost the same as the monomer's $\beta$-value. 
$\beta$ for porous aggregates obtained by Mie Theory with Maxwell-Garnett approximation for effective dielectric function are also plotted ({\it green dashed lines}).
When $s_0=0.1\mu$m, the wavelength of stellar radiation is shorter than the monomer radius and the  $\beta$-value obtained by effective medium theory deviates from that of T-Matrix Method.}
\label{fig:beta_clr}
\end{center}
\end{figure}

We calculate $\beta$-values for porous aggregates and compact grains based on Eq.(\ref{eq:beta}) and Eq.(\ref{eq:qrp}) and plot them in Figure \ref{fig:beta_clr}.
For comparison, $\beta$ for porous aggregates with $s_0=0.01$ $\mu$m is plotted in addition to our fiducial model with $s_0=0.1$ $\mu$m.
In Table \ref{tab:rc}, we summarize the sizes of aggregates composed of $0.1$ $\mu$m-sized monomers and their $\beta$-values.
The absorption and scattering cross sections are calculated using the Mie Theory for compact grains, and using the T-Matrix Method 
for porous aggregates.
Properties of light scattering and absorption by porous aggregates will be discussed in \citet{tazaki_inprep}.
Hence, we only describe $\beta$-value in this paper. 
In the case of compact grain, as the size increases, $\beta$-value also increases when the grain size is small.
When $x\sim 1$, $\beta$-curve strongly depends on optical constants due to resonance and interference effect \citep[see e.g., Appendix B in][]{miyake93}.
In geometrical optics regime where $x\gg1$ and a monomer is optically thick, 
$Q_{\rm{RP}}$ does not depend on size parameter and $\beta$-value for the compact grain only depends on
 area-to-mass ratio, that is, $\beta\propto{s}^{-1}$.
Therefore as the size increases, $\beta$-value for compact grain decreases. 
In contrast, in the case of porous aggregates $\beta$-value remains high even if the monomer number or characteristic radius increases \citep[e.g.,][]{kimura02, kohler07}. 
As we mentioned in Sect.2.3.1, if $x\ll{1}$, absorption dominate the radiation pressure efficiency, $Q_{\rm RP} \sim Q_{\rm abs}$.
In addition, in this regime absorption cross-section of BCCA can be simply described  by superposition of monomer's absorption cross-section, $C_{\rm abs}(N)=NC_{\rm abs}(N=1)$ \citep[e.g.,][]{kolokolova07, kataoka14}. Therefore, $\beta_{\rm N,BCCA}/\beta_{\rm N=1}\propto{N}^0$.

In Figure \ref{fig:beta_clr}, we also plot $\beta$-value obtained using effective medium theory with Maxwell-Garnet mixing rules
 \citep{chylek00} with filling factor defined by Eq. (\ref{eq:filling}) for comparison with T-Matrix Method.
 In the case of $s_0=0.01\mu$m, the result of effective medium theory 
does not deviate from that of T-Matrix Method very much, although it does if $s_0=0.1$ $\mu$m.
This is because, in the latter case, the size parameter of a monomer approaches to unity, then 2nd or higher order of Lorentz-Mie coefficients cannot be negligible and the EMT approximation breaks down.

\subsection{Outward Drift of Grains in Surface Layer}
\begin{figure*}[htbp]
\begin{center}
\includegraphics[height=6.2cm]{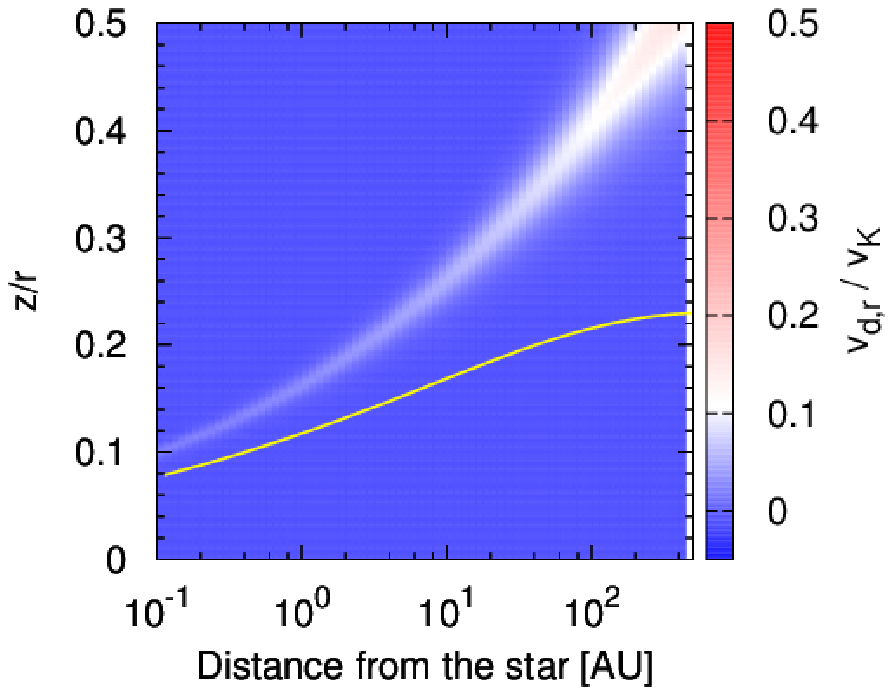}
\includegraphics[height=6.2cm]{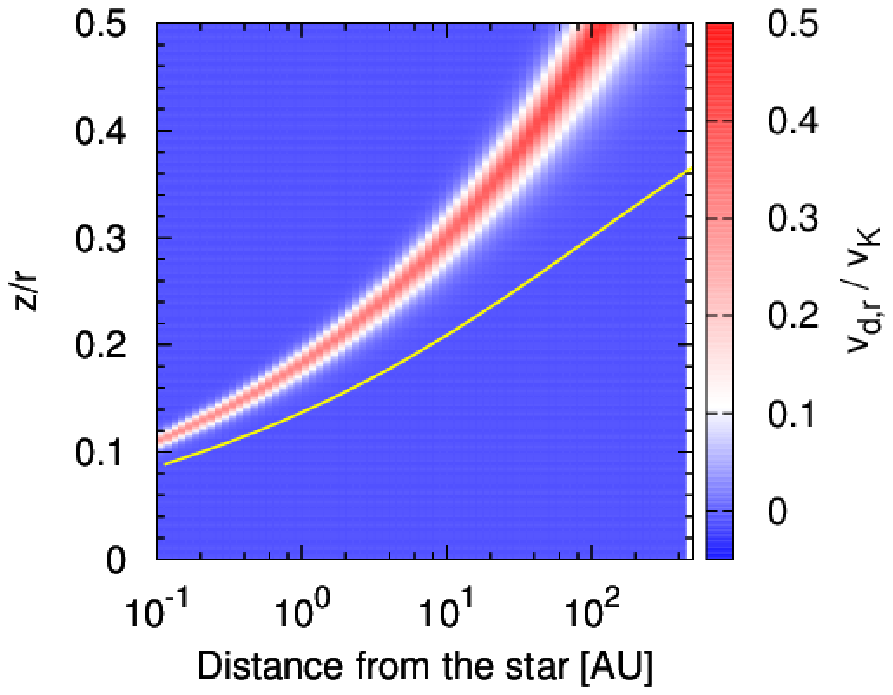}
\caption{The blowing out region at the surface layer of protoplanetary disks. Yellow solid line indicates the height where $\tau=1$ defined in Eq. (\ref{eq:tau}). Top panel is for the compact grains and bottom panel is for the porous aggregates. The grain radius of 4.37 $\mu$m is adopted for both models.
The color scale represents the radial velocity of dust particle normalized by Keplerian velocity.}
\label{fig:veldr}
\end{center}
\end{figure*}

Figure \ref{fig:veldr} shows the radial velocity of dust grain obtained from Eq. ({\ref{eq:vel_dr}}) .
Above the photosphere ($\tau=1$), dust grains are exposed to stellar radiation pressure and they move along spiral-out orbit. 
Outward drift velocity becomes maximum where non-dimensional stopping time equals to unity. In this case, $v_{d,r}|_{T_s=1}\sim\beta{v_K}/2$ when $\beta\gg\eta$. $\beta$ for compact grain with $s=4.37$ $\mu$m (equivalent to the characteristic radius of BCCA with 1024 monomers) is $\beta=0.043$,
 whereas $\beta=0.62$ for porous aggregates.
Therefore, the outward drift velocity for
porous aggregates is 62\% of the Keplerian velocity while 4.3\% for compact grains.
However, the velocity with which most of mass is transported is relatively slower than this velocity because, just above the photosphere, dust grains are weakly coupled to gas, that is, $T_s<1$ (see Figure \ref{fig:veldr}). Since the dust density rapidly decreases as the height increases, the outward mass flux is determined by the radial velocity and dust density at illuminated surface, or just above $\tau=1$.
At the surface layer, the radial velocity can be described approximately as $v_{d,r}(r,z_{\rm sur})=v_{g,r}+\beta{v_K}T_s$ where we neglect $\eta$ since $\eta\ll\beta$ and omit the second order of $T_s$. 

\begin{figure}[b]
\begin{center}
\includegraphics[height=6.5cm,keepaspectratio]{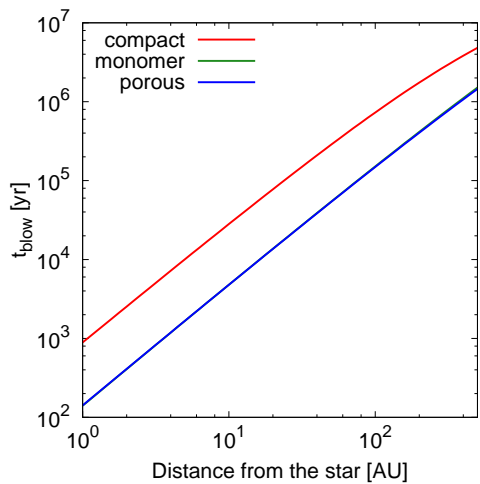}
\caption{Blowing out timescales for 
the compact grain with $s=4.37\mu$m ({\it red line}), BCCA with N=1024 and $s_0=0.1\mu$m ({\it green line}) and monomer with $s_0=0.1\mu$m ({\it blue line}).}
\label{fig:tb}
\end{center}
\end{figure}

We define blowing out timescale as $t_b=r/|v_{d,r,z_{sur}}|$ and plot it in Figure \ref{fig:tb}. $z_{sur}$ is the height where $\tau=1$. 
Figure \ref{fig:tb} also shows that the monomer or porous aggregates blow outward with shorter timescale than compact grain,
 since $\beta$ of porous aggregates is larger than that of compact grain by an order of magnitude. 
It is also inferred that the porous aggregates cannot move beyond around $200$ AU before dispersal of a disk if we consider that disk dispersal timescale is an order of Myr \citep[e.g.,][]{hernandez08}. 
Note that since stopping time of porous aggregates and compact grain is different, the height of illuminated surface is also different between them. 
Thus, the difference in blowing out timescales of compact grain and porous aggregates comes from both $\beta$-value and $T_s$.

Figure \ref{fig:tb} implies that the outward flow velocity at the surface layer is high. However the outward mass flux
is not very large since only small fraction of dust grains can reside in the optically thin surface layer.
We evaluate the mass flux by surface outward flow defined by
\begin{eqnarray}
F_{\rm{out}}(r) = 2 \int_{z_{sur}}^{\infty} 2\pi{r}\rho_d(r,z)|v_{d,r}(r,z)|dz,
\end{eqnarray}
where the factor 2 arises from the both sides of disk. 
The outward mass flux can be written approximately as $F_{\rm{out}}\sim 2\pi{r}|v_{d,r}(r,z_{sur})|\Sigma_{d,sur}$
 where $\Sigma_{d,sur}$ is
\begin{equation}
\Sigma_{d,sur}=2 \int_{z_{sur}}^{\infty} \rho_d(r,z)dz.
\end{equation}

\begin{figure}[t]
\begin{center}
\includegraphics[height=6.5cm,keepaspectratio]{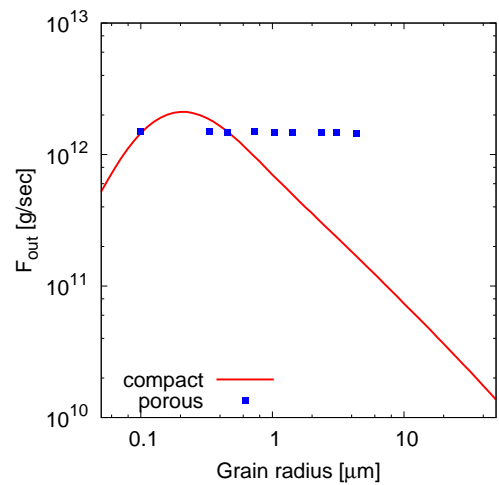}
\caption{The outward mass flux at $r=1$ AU is plotted as a function of particle radius. Red line represents the mass flux for the compact grain and blue squares are for porous aggregates with $N=1 \sim1024$ and $s_0=0.1\mu$m.} 
\label{fig:mf}
\end{center}
\end{figure}

In Figure \ref{fig:mf}, we plot the outward mass flux at $r=1$AU as a function of grain radius. 
Figure \ref{fig:mf} is qualitatively similar to Figure \ref{fig:beta_clr}.
Outward mass flux is $1.4\times10^{12}$ g/sec for porous aggregates with $N=1024$ whereas it is $1.7\times10^{11}$ g/sec for compact grain with the same radius. 
Thus, porous aggregates are blown out more effectively than compact grain by
an order of magnitude. 
What we should mention here is the porous aggregation itself does not affect the outward mass flux of dust grain. Even if they grow to be larger size, porous aggregates can move outward as effectively as a monomer whereas compact grains cannot since the radiation pressure becomes small.
$\beta$ of porous aggregates keeps constant as they grow
until their fractal dimension deviates from 2 owing to compaction such as by collisional compression \citep[e.g.,][]{okuzumi12}.

\citet{juhasz12} reported
crystalline features are detected from inner region of PPDs soon after the outburst, however these crystalline features decrease as time goes on.
They construct a model and conclude that depletion of the crystalline grains after the outburst cannot be explained by the vertical mixing and they are transported to large disk radii due to, for example, radiation pressure.

\subsection{Timescales of Vertical Dust Dynamics}

In the previous section, we do not take into account vertical motion of dust grains.
In PPDs, dust grains can also move in the vertical direction due to stellar gravity and turbulent mixing.
Once grains sink into optically thick region, they cannot be exposed to the radiation pressure any more. 
Hence, it is valuable to mention here about timescales of vertical motion and radial motion.

For simplicity, only settling and turbulent stirring are considered as the vertical motion of dust grains.
Dust grains settle down to the disk midplane due to the stellar gravity and experience the gas drag force. Thus, the settling velocity $v_{\rm{sett}}$ can be determined by the balance between the stellar gravity and the gas drag force, and the settling timescale can be described by
\begin{eqnarray}
t_{\rm{sett}}=\frac{z}{v_{\rm{sett}}}=\frac{4}{3}\frac{\sigma}{m}\frac{\rho_gv_t}{\Omega_{K, {\rm mid}}^2}, 
\end{eqnarray}
\citep{dullemond04}. Since the gas density decreases as disk height increases, the settling timescale becomes small at the large height of the disk. 

\begin{figure*}[htbp]
\begin{center}
\includegraphics[height=6.2cm,keepaspectratio]{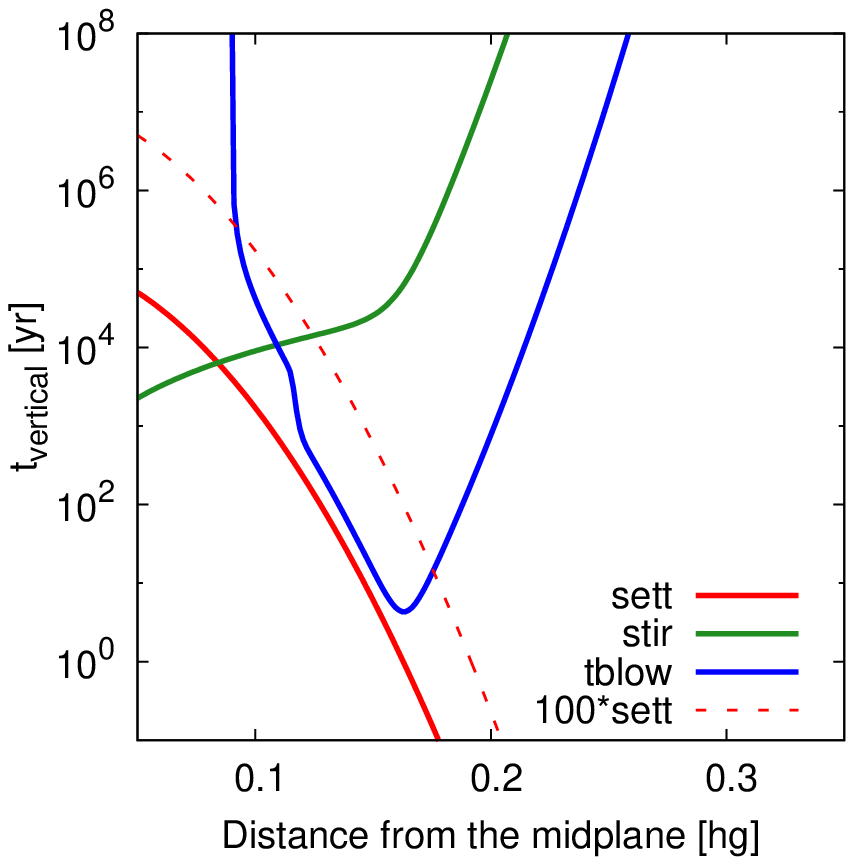}
\includegraphics[height=6.2cm,keepaspectratio]{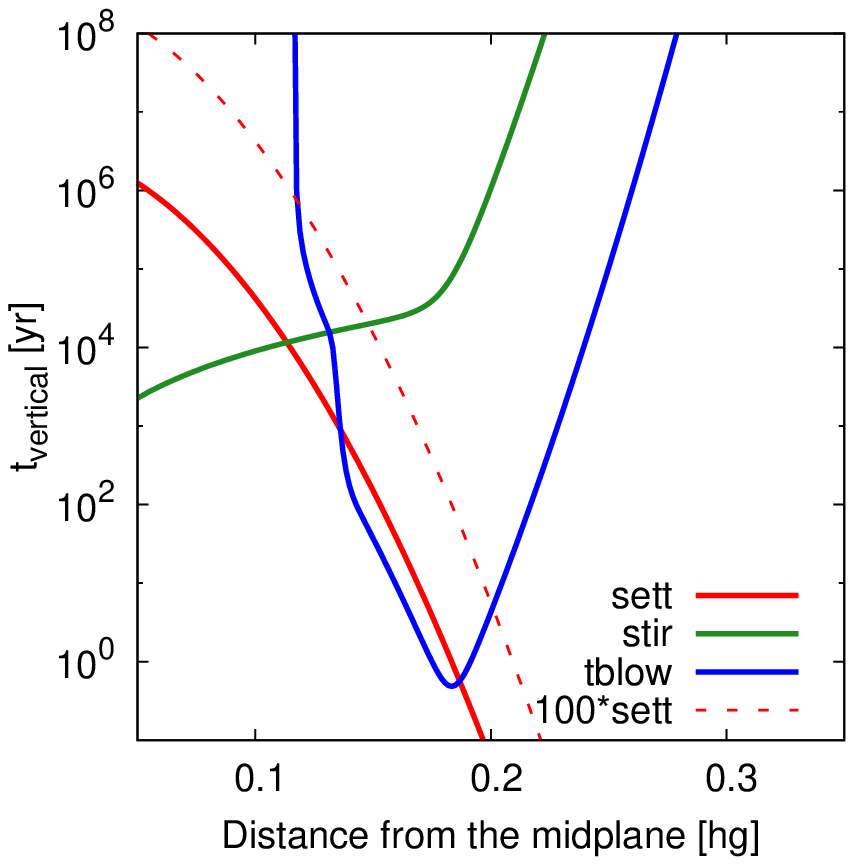}
\caption{
Blowing timescale $t_b$ ({\it blue lines}), stirring timescale $t_{stir}$ ({\it green lines}) and settling timescale $t_{sett}$ ({\it red lines}) at $r=1$AU.
Settling timescale multiplied by 100 is also plotted with red dotted lines.
The result for compact grains and porous aggregates are shown in top and bottom panels, respectively.}
\label{fig:t_vert}
\end{center}
\end{figure*}

Meanwhile, dust grains are stirred up by turbulent gas. The timescale of turbulent diffusion can be written in the form of $t_{\rm{stir}}=z^2/D_{\rm{diff}}$, where $D_{\rm{diff}}$ is the diffusion coefficient of dust grains and depends on (i) the strength of turbulence and (ii) the degree of coupling between gas and dust. Since the strong turbulence yields the large diffusion coefficient, the diffusion coefficient is set to be proportional to the turbulent viscosity of gas, $\nu$. On the other hand, if dust grains are completely decoupled from gas, then the diffusion coefficient must be zero. Thus, we get $D_{\rm{diff}} = \nu/Sc$ where $Sc$ is the Schmidt number defined by $Sc = 1 + St$ \citep{cuzzi93}, and $St$ is the Stokes number defined by $St \equiv (v_t/c_s)T_s$. For the well coupled case, the Schmidt number approaches unity, and for the completely decoupled case, this number goes to infinity. If we adopt the $\alpha$ model for the turbulent viscosity (Eq.(\ref{eq:vis})),
we get the stirring timescale,
\begin{equation}
t_{\rm{stir}}=\frac{Sc}{\alpha\Omega_{K, {\rm mid}}}\frac{z^2}{h_{g}^2} .
\end{equation}
We compare the blow out timescale and the timescale of vertical dynamics (settling and stirring).
Figure \ref{fig:t_vert} shows vertical timescales of porous aggregates and compact grains.
The depletion latitude of dust grains can be determined by the balance between $t_{\rm{stir}}$ and $100t_{\rm{sett}}$ \citep{dullemond04}
 and $100t_{\rm{sett}}$ is also plotted with dashed line in Figure \ref{fig:t_vert}.
In the case of porous aggregates, the blowing out timescale is shorter than vertical dynamical timescales.
Therefore porous aggregates can be blown out ballistically.
In contrast, compact grains experience faster settling due to their loose coupling to gas. Thus they settle down to midplane and 
can not drift outward anymore.
Note that \citet{vinkovic09} suggested that the grain is lifted up by radiation pressure force from disk emission, hence
 the net effect of settling is reduced. Under such circumstances, grain might be transported without sink to disk midplane.
 To verify this effect, further calculation is necessary. 
 
Outward mass flux for porous aggregates with $s_0=0.1 \mu$m and $N=1024$ at $1$ AU is $1.4\times10^{12}$ g/sec.
Porous aggregates are expected to avoid fast settling, then, optimistically, total mass transported from inner region to comet
 formation region is at most $1.4\times10^{12}$ g/sec $\times {\rm Myr} \sim 4.5\times10^{25}$ g, which is comparable to the mass of the Kuiper Belt Objects \citep[e.g.,][]{chiang07}.

\section{Implications to the origin of cometary grains}
Our results suggest that porous aggregates like CP-IDPs are blown out efficiently, whereas compact grain like calcium aluminum rich inclusions (CAIs) are hardly blown out by radiation pressure in the surface layer of the protosolar nebula.
It is worth noting that this result does not always contradict with the result of {\it Stardust} mission, which  is a sample return mission from comet 81P/Wild 2 conducted by NASA. 
{\it Stardust} samples showed that comet 81P/Wild 2 contains not only crystalline silicate but also refractory grains such as CAIs and chondrule fragments
 that are usually found in meteorites \citep[][and references therein]{brownlee14}. 
 However, crystallographically, crystalline silicates found in collected {\it stardust} samples and CP-IDPs have different properties. 
 For instance the enstatite whiskers in CP-IDPs tend to elongate in the [100]-direction whereas
enstatite in {\it Stardust} samples shows structure in the [001]-direction that is, when not equiaxial, commonly found in meteorites \citep{ishii08}.
Since materials in comet 81P/Wild 2 resemble to that in meteorites originating from asteroids, these results imply that in the early solar nebula
comet 81P/Wild 2 could originally be a member of asteroids. Indeed, recent N-body calculation suggested planetesimals could be
 scattered from a few AU to far beyond current Neptunian orbit \citep[e.g.,][]{nagasawa14}. 
Therefore, if this were true, the origin of crystalline silicate in comet 81P/Wild 2 and CP-IDPs are not necessary the same so that
the large compact grains, like CAIs, are not necessary to transport from inner disk to outer disk by stellar radiation pressure. 

\section{summary}
We have studied the surface outflow of dust grains by the stellar radiation pressure 
to explain the presence of crystalline silicate in comets.
Especially, we took into account the porosity of dust grain to mimic the CP-IDPs
which originate from comets.
First, we confirmed the radiation pressure for the porous aggregates is determined
 by the monomer's properties.
Based on this results, we have calculated the outward mass flux at the surface layer.
As a results, porous aggregates show much higher mass flux compared to the compact grains 
when their radii are equivalent.
This suggests porous aggregates like CP-IDPs are transported to outer region of PPDs efficiently, 
thus our model could be a possible candidate for the transport mechanism of 
crystalline silicates in porous aggregates from inner hot regions to outer cometary regions.

\acknowledgments
R.T. thanks to Yasuhiko Okada for technical advice about T-Matrix Method.
R.T. also would like to appreciate 
useful discussions with Hiroshi Kimura and Junya Matsuno.
The numerical calculations were carried out on SR16000 at  YITP in Kyoto University.
This work is supported by Grant-in-Aids for Scientific Research, 23103005 and 25400229.

\end{document}